# Single-photon polarization tomography with an integrated metal-superconductor nanowire array

**Pierre Brosseau[1,2], Jiawei Wang[1,2], Giorgio Adamo[1,2,3], Anton N. Vetlugin[1,2], Cesare Soci[1,2,4,*]**

*1. Centre for Disruptive Photonic Technologies, The Photonics Institute, Nanyang Technological University, 637371, Singapore*

*2. Division of Physics and Applied Physics, School of Physical and Mathematical Sciences, Nanyang Technological University, 637371, Singapore*

*3. Optoelectronics Research Centre, University of Southampton, Highfield, Southampton SO17 1BJ, UK*

*4. School of Electrical and Electronic Engineering, Nanyang Technological University, 639798, Singapore*

**csoci@ntu.edu.sg*

## Abstract

Light polarization is a primary degree of freedom for encoding quantum information. The scaling up of photonic quantum networks and computer architecture depends crucially on its precise characterization. This is typically achieved by placing external waveplates, polarizers, moving mounts, and recently metasurfaces, on top of the detectors. All these solutions complicate integration and scaling. Here we break convention with traditional architecture and present a monolithic, self-aligned metal-superconductor nanowire single photon detector (M-SNSPD) possessing intrinsic full polarization selectivity. Gold nanowires, co-fabricated atop NbTiN superconducting nanowires within the same lithographic footprint, act as polarization-selective plasmonic metamaterials inducing resonant absorption in the NbTiN. U-shaped wires provide linear polarization selectivity, while S-shaped meanders distinguish circular polarization, while retaining the high-count rates and low dark count rates of conventional SNSPDs. By arranging them into a four-pixel array we realize simultaneous projection onto four polarizations and demonstrate continuous polarization state tomography with an ensemble average fidelity exceeding 98%. Our approach opens new avenues towards scalable detector arrays with integrated plasmonic functionalities, for single photon polarimetry, imaging and spectroscopy.

## Main

Accurate characterization of the polarization state of single photons is crucial to many photonic quantum technologies, including the characterization of entangled-photon sources [1], the interfacing with quantum memories [2], the benchmarking of photonic quantum gates [3] and polarization-encoded quantum communications [4]. All single photon detection technologies currently lack the intrinsic capability of polarization state discrimination. Even superconducting nanowire single photon detectors (SNSPDs) [5], arguably the best performing platform due to their high detection efficiency [6], very broad bandwidth operability, low timing jitter and ultralow dark-count rates, are only limited to linear polarimetry leveraging the absorption anisotropy brought about by the geometry of the nanowire design [7-11]. While this represents a useful handle on the polarization degree of freedom, it is still incomplete. Consequently, single photon polarization analysis is currently performed by cascading arrangements of waveplates, beamsplitters and polarizers before a detector [12]. Although effective, this approach inherently relies on external bulky and mechanically reconfigurable optics, which limit scalability and are incompatible with the emerging integrated quantum photonic technologies.

More recently, planar metamaterials, thin two-dimensional periodic arrangements of nanoscale optical resonators with extreme ability of controlling light-matter coupling at the nanoscale on-demand, have been used to improve absorption [13,14]. Plasmonic metamaterials are of particular interest for these applications because they localize and enhance optical fields outside the metallic nanoscale resonators, in proximity to the absorbing layers [15]. All current metamaterials solutions, where the plasmonic resonators were added on top of stand-alone single photon detectors, including SNSPDs, have focused on increasing absorption and improving overall efficiency [16,17], disregarding full polarization selectivity or access to a complete polarization measurement basis.
In this work we embed the single photon detection ability of superconducting nanowires and the near field optical mode engineering capability of plasmonic metamaterials into a single workflow and co-design a self-aligned metal-superconductor nanowire single photon detecting device (M-SNSPD), as illustrated in Fig. 1, where a four-pixel array is shown simultaneously projecting an incident state $\rho$ onto four distinct polarization bases mapped to the Poincaré sphere.

A gold nanowire is fabricated directly atop a NbTiN superconducting nanowire using the same electron-beam lithography step, so that both layers share an identical footprint. After lift-off, the gold nanowire serves as a hard etch mask for the NbTiN layer beneath, defining both wires in a single patterning step on a SiO2-on-distributed-Bragg-reflector substrate. Rather than directing light onto a passive absorber, the gold overlayer shapes the electromagnetic near field at the point of absorption, its geometry, shape, symmetry and chirality determine which polarization state couples efficiently into the superconductor and triggers a detection event.

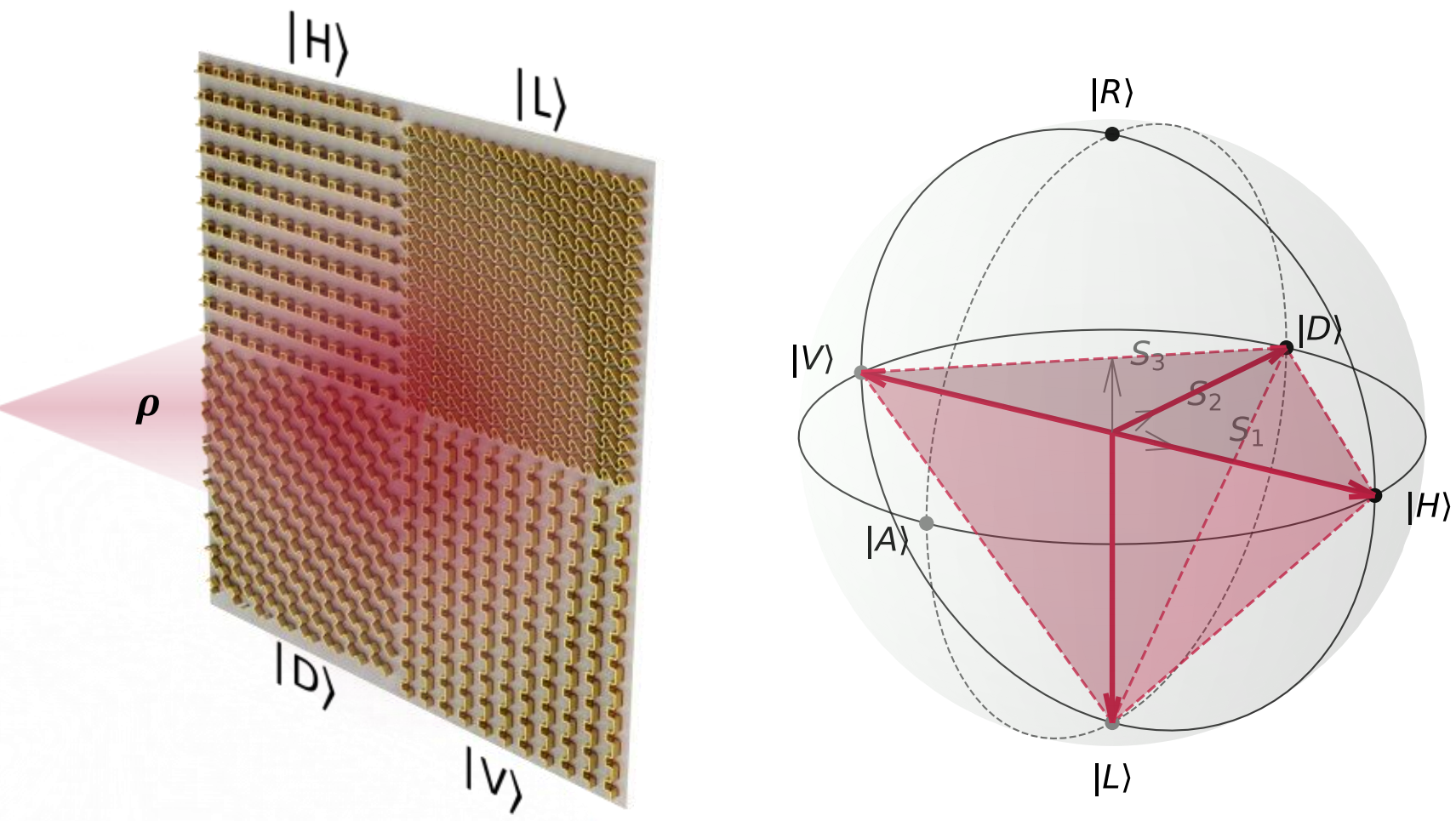

**Fig. 1. Monolithic metal-superconductor nanowire single photon detector array for polarization state tomography.** An incident single photon state $\rho$ illuminates a four-pixel M-SNSPD array, where each pixel implements a distinct polarization projection. Three pixels comprise U-shaped gold nanowires oriented to project onto |H⟩, |V⟩, and |D⟩ while a fourth pixel with an S-shaped chiral meander projects onto |L⟩. Each pixel is mapped to its corresponding point on the Poincaré sphere, with the Stokes axes $S_1$, $S_2$, $S_3$ indicated. Together, the four projections form an informationally complete measurement basis for reconstruction of an arbitrary polarization qubit, without sequential measurements, moving optics, or any external polarizing element.

Two unit-cell geometries implement the polarization projections needed for complete state reconstruction. A U-shaped gold nanowire has structural anisotropy along one axis; it concentrates the near field for a linear polarization aligned with its open end while leaving the orthogonal polarization largely uncoupled. Rotating the U-cell rotates the preferred axis. A chiral S-shaped meander lacks mirror symmetry and generates a handed near field distribution that couples selectively to one circular polarization handedness.

## Results

### Geometry-engineered polarization selectivity

For circular polarization, we break in-plane mirror symmetry using a chiral S-shaped meander (Fig. 2a). This nanoscale chirality induces a helicity-dependent electromagnetic response, resulting in differential absorption between left and right circularly polarized photons [18-22]. Insights into the modal coupling are provided by three-dimensional finite difference time domain (FDTD) simulations. The S-shaped device shows differential absorption between helicities, reaching 15% for |L⟩ compared to 5% for |R⟩ at 1550nm, achieving a circular polarization visibility of ≈0.5 (Fig. 2b). This simulated absorption asymmetry is directly reflected in the polarization-resolved reflection spectra (Fig. 2c). The measured curves for |R⟩ and |L⟩ diverge around 1550nm, with the |L⟩ reflection dipping significantly below |R⟩, consistent with preferential absorption of the left-handed component. The measured and simulated spectra show close agreement across the telecom C-band confirming that the FDTD model accurately captures the plasmonic near-field coupling mechanism responsible for the chiral response.

To resolve the linear basis, we use a periodic array of U-shaped nanowires (Fig. 2d). The structural anisotropy of the U-motif leads to preferential coupling of light polarized parallel to the wire segments [23], while the orthogonal component remains largely uncoupled. For this geometry, we observe a pronounced divergence between horizontal |H⟩ and vertical |V⟩ states, with a peak absorption of 44% at 1550nm for the preferred axis, yielding a linear-polarization visibility of ≈0.4 (Fig. 2e). The reflection spectra (Fig. 2f) confirm this behavior. |H⟩ and |V⟩ exhibit a clear and sustained separation across the telecom band, with the |H⟩ reflection remaining consistently lower than |V⟩ near 1550nm, indicating stronger absorption for the preferred polarization axis. The measured and simulated spectra are in good agreement, with minor deviations attributable to fabrication imperfections such as edge roughness and dimensional tolerances in the nanowire geometry.

Devices are fabricated via a multi-step nanofabrication process, starting with the deposition of thin-film NbTiN onto a SiO2-on-DBR substrate via DC magnetron sputtering [24]. Nanowires are patterned using electron beam lithography, followed by thermal evaporation and lift-off to form gold nanowires. These metallic wires serve as a hard mask during a subsequent reactive ion etch, which transfers the geometry into the NbTiN layer to produce self-aligned metal-superconductor nanowires.

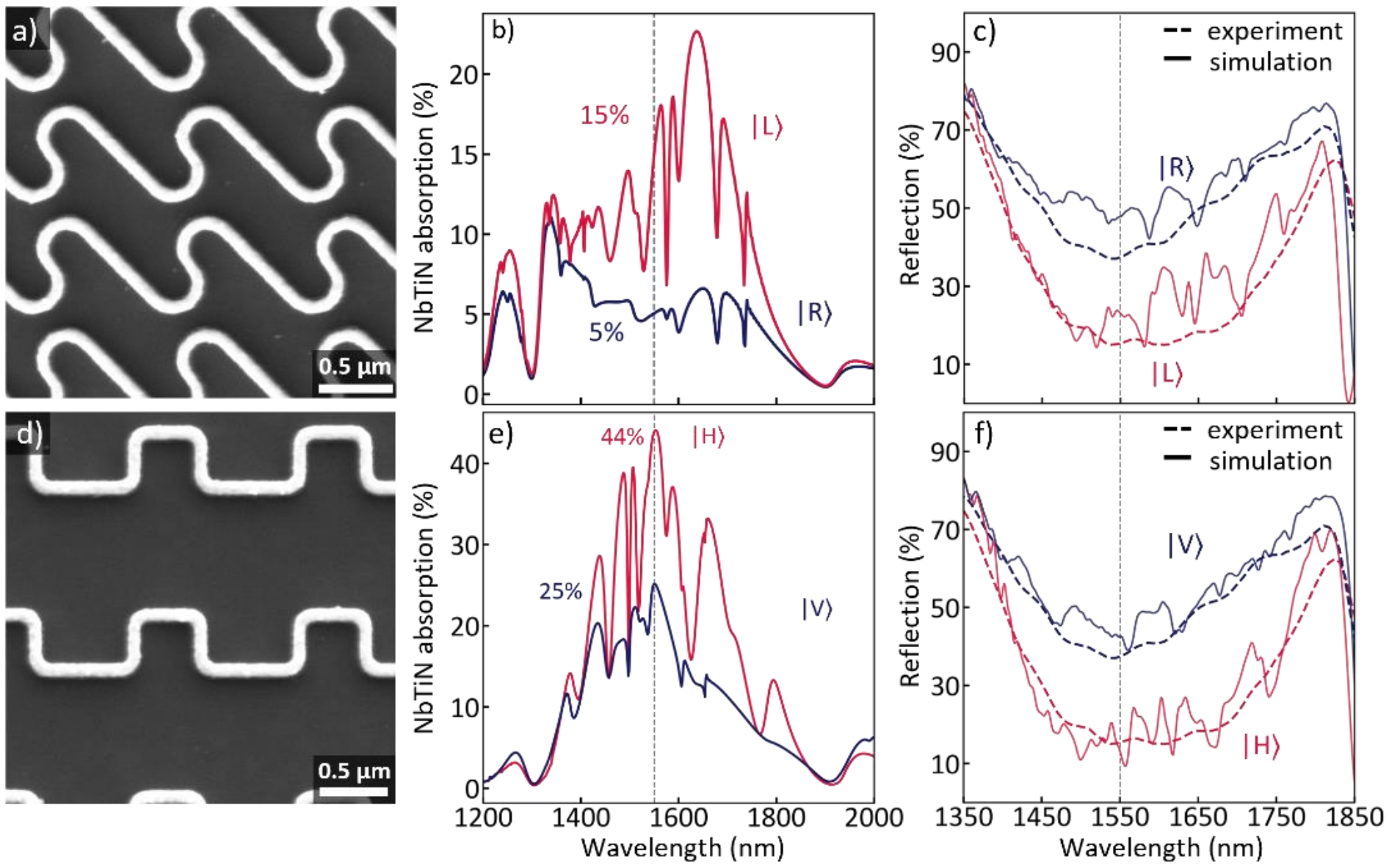

**Fig. 2. Geometry-engineered polarization selectivity.** (a) SEM image of the chiral S-shaped nanowires enabling circular polarization discrimination. (b) Simulated NbTiN absorption for right and left-handed circular polarizations. (c) Measured (dashed line) and simulated (solid line) reflection spectra confirming helicity-dependent optical response. (d) SEM image of the U-shaped nanowires designed for linear polarization sensitivity. (e) Simulated NbTiN absorption for horizontal and vertical polarizations. (f) Measured (dashed line) and simulated (solid line) reflection spectra showing polarization-dependent response.

## Single photon counting performance

The electrical dynamics of the hybrid detector are defined by the equivalent circuit shown in Fig. 3a (inset). The superconducting nanowire acts as a kinetic inductance $L_k$ in series with a time-dependent resistance [25]. The metallic nanowire forms a parallel branch with resistance $R_m$ and inductance $L_m$. At 3 K, we measure $R_m = 4.1\ k\Omega$. Because the typical peak hotspot resistance in NbTiN SNSPDs is also on the order of a few kilo-ohms, this metallic branch provides an alternative pathway for transient current redistribution during hotspot formation, while remaining electrically passive during steady state biasing. We compared the hybrid device to a reference bare superconducting nanowire fabricated under identical conditions. Both exhibit nearly identical electrical transients (Fig. 3a), with peak amplitudes of $\sim$100mV, rise time <1ns, and exponential decay constants of 6-7 ns. The unchanged decay dynamics indicate that $L_m \ll L_k$ and that the metallic branch does not distort the reset process. Furthermore, the dark count rate (DCR) behavior is preserved (Fig. 3b). Time traces recorded at $0.6I_c$ and $0.8I_c$ show comparable DCR for both the M-SNSPD ($\mu = 7.12 \pm 3.4$ counts) and the reference SNSPD ($\mu = 9.60 \pm 3.8$ counts), demonstrating that the metallic overlayer introduces no measurable excess dissipation or thermal hotspots.

To verify polarization-resolved single photon response, we illuminate each pixel through a free-space collimator followed by a rotating waveplate. For the U-shaped linear pixel (Fig. 3c), a rotating half-wave plate (HWP) continuously varies the output linear polarization angle, and the count rate follows Malus's law $R(\theta) = A + C\ cos^2(\pi/P \cdot (\theta - \phi_0))$, with $A = 36.8$kHz, $C = 27.1$kHz, $P = 90.0°$, yielding a visibility of $\approx 0.27$. The count rate maxima and minima coincide with |H⟩ and |V⟩, confirming a projective measurement onto the geometrically defined linear polarization basis. For the chiral S-shaped pixel (Fig. 3d), a rotating quarter-wave plate (QWP) converts the input linear polarization into a continuously varying sequence of elliptical states. The detector response encodes the full polarization sensitivity of the pixel, including circular dichroism, and is well described by $R(\theta) = A + B\cos(4\theta + \phi_4) + C\sin(2\theta + \phi_2)$, with $A = 39.7$kHz, $B = 3.2$kHz, $C = 19.3$kHz. The large $\sin(2\theta)$amplitude relative to the $\cos(4\theta)$ term yields a circularity of $|C|/(|B| + |C|) \approx 0.86$, confirming dominant circular dichroism and a strong preferential response to |L⟩ over |R⟩.

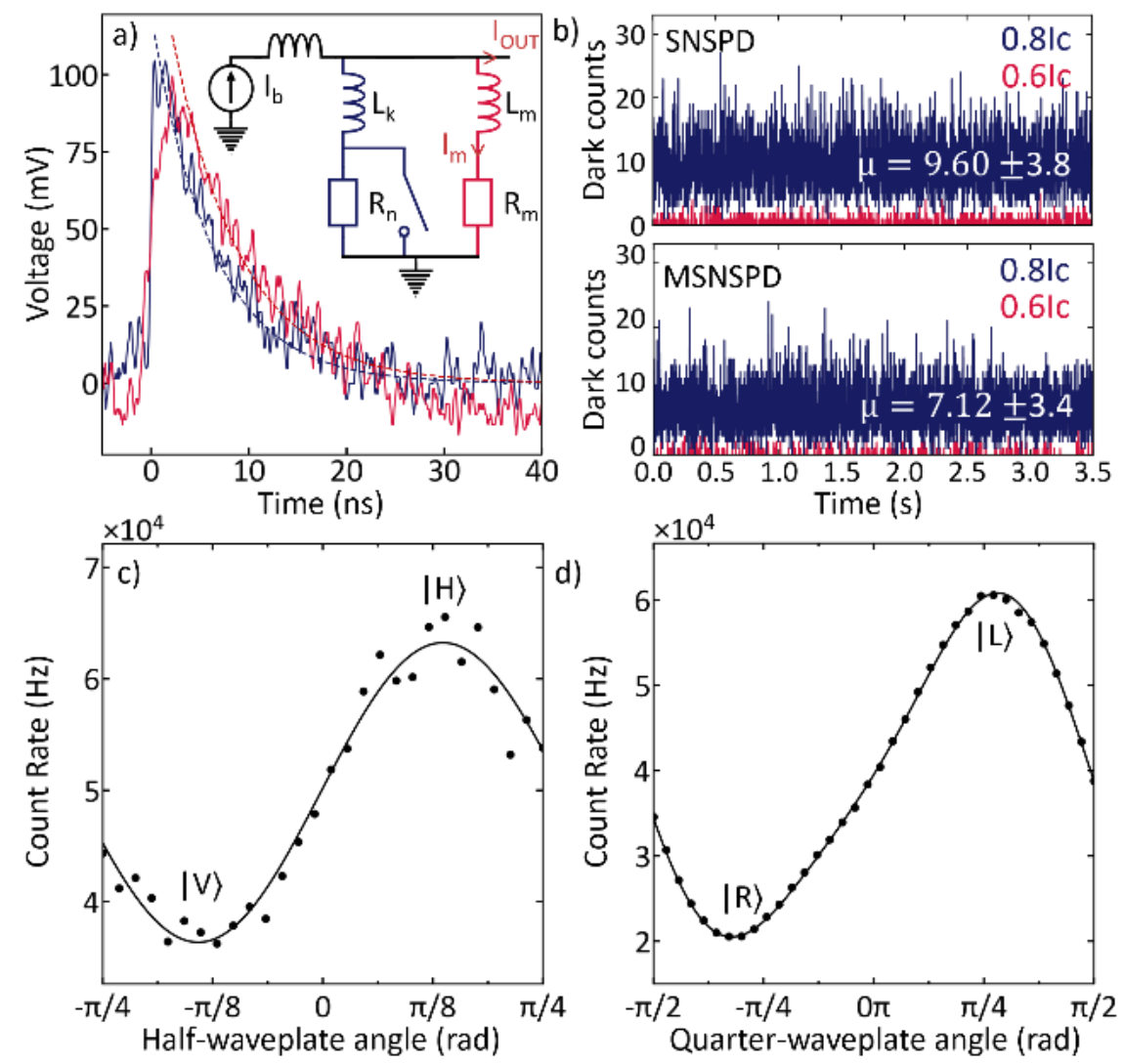

**Fig. 3. Electrical performance and polarization-resolved single photon response of hybrid detectors.** (a) Electrical pulse traces measured from the reference SNSPD (blue) and M-SNSPD (red). The inset shows the equivalent circuit model including the kinetic inductance $L_k$ and the metallic branch ($R_m, L_m$). (b) Dark count time traces at $0.6I_c$ and $0.8I_c$ for the M-SNSPD (top) and reference SNSPD (bottom), showing comparable noise performance. (c) Photon count rate versus HWP angle for the U-shaped linear pixel, fitted with Malus's law (solid line); count rate extrema correspond to |H⟩ and |V⟩. (d) Count rate versus QWP angle for the S-shaped chiral pixel, fitted with a three-component model (solid line); the dominant $\sin(2\theta)$ term confirms intrinsic circular dichroism.

### Polarization state tomography

Complete characterization of a polarization qubit requires simultaneous projection onto an informationally complete set of polarization states [12], a task conventionally performed by rotating a waveplate through several angles in front of a fixed linear polarizer and single detector, which is slow, susceptible to mechanical drift, and incompatible with integrated photonic circuits [26]. The four-pixel array demonstrated here acquires all four projections in parallel and continuously, removing the need for any moving optical element. A scanning electron micrograph of the array is shown in Fig. 4a. Each pixel targets one of the states |H⟩, |V⟩, |D⟩ or |L⟩, whose measurement axes are distributed across the Poincaré sphere with a condition number $\kappa \approx 3.23$.

Prior to reconstructing unknown states, the instrument matrix of the array is calibrated using an overcomplete set of six known input polarizations (|H⟩, |V⟩, |D⟩, |A⟩, |R⟩, |L⟩). Dark counts are subtracted and count rates are normalized to the input photon flux. For each unknown input, the four measured count rates are mapped to a physical Stokes vector by constrained maximum likelihood inversion of the calibrated matrix, from which the density matrix is directly constructed.

We test the array on nineteen input polarization states distributed across the Poincaré sphere, accumulating counts continuously in all four pixels simultaneously. Reconstruction fidelity is evaluated after each integration window as additional photons are added (Fig. 4b). At short integration times, the individual fidelity traces (light blue) fluctuate widely, reflecting the broad likelihood landscape when few photons have been detected. As counts accumulate, all reconstructions converge toward the true state. The ensemble average (dark blue) exceeds 98% fidelity, and the Bloch sphere inset illustrates this convergence geometrically; the 10ms estimate (dashed arrow) deviates substantially from the true state (black dot), while the 100ms reconstruction (red arrow) closely overlaps it. Convergence is faster for states aligned with the more informative pixel axes and slower for states that fall near the weakest measurement direction, a direct consequence of the anisotropic Fisher information distribution across the four pixels.

A representative reconstruction after 15ms (Fig. 4c) demonstrates the quality of the result at the single-state level. The real and imaginary parts of the reconstructed density matrix $\rho_{out}$ closely match those of the input state $\rho_{in}$ in both amplitude and phase, with residual deviations consistent with the statistical uncertainty expected from the measured Fisher information. The resulting fidelity of 99.57% confirms that near-unity polarization-state tomography is achievable at the millisecond timescale without any external polarizing optics.

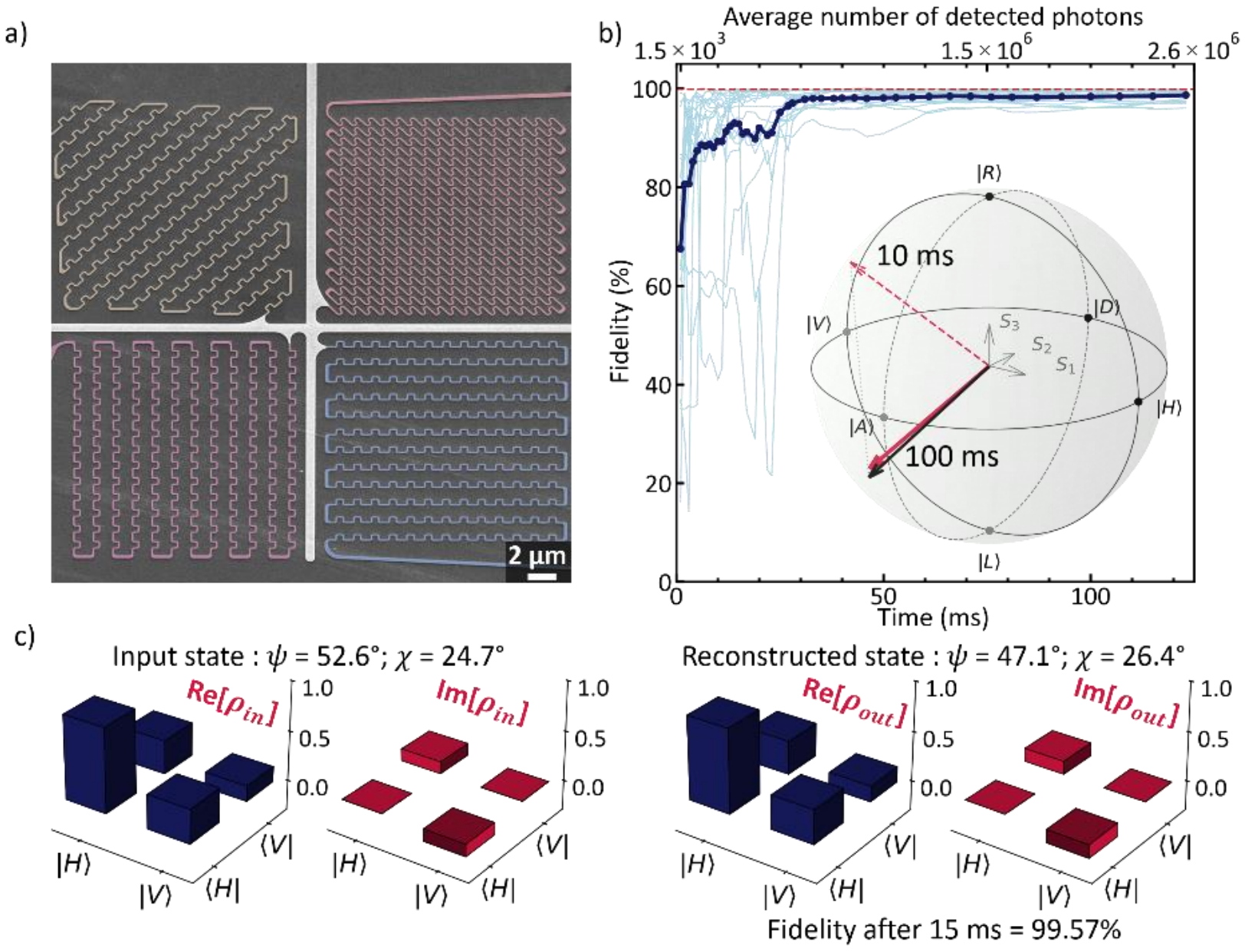

**Fig. 4. Experimental demonstration of continuous, spatially multiplexed quantum state tomography.** (a) Scanning electron micrograph of the polarization selective four-pixel array. (b) Experimental fidelity convergence for 19 input states, achieving an ensemble average >98% within 100ms; inset shows state convergence on the Poincaré sphere at 10 and 100ms. (c) Representative high-fidelity (99.57%) density matrix reconstruction after 15ms of integration, showing real and imaginary parts of $\rho_{in}$ (left) and $\rho_{out}$ (right).

To frame the experimental operating conditions, we model the expected reconstruction fidelity as a function of photon number and pixel visibility (Fig. 5a). Three regimes emerge. At low photon counts the reconstruction is shot noise dominated; fidelity is largely insensitive to visibility and improves only with accumulated counts. At intermediate counts, fidelity depends jointly on photon number and measurement contrast, with iso-fidelity contours scaling as $N \propto 1/V^2$, so that a reduction in visibility imposes a quadratic penalty on integration time. At high photon counts, the reconstruction saturates, limited in practice by systematic calibration errors and residual POVM overlap. The full analytical framework, including Bayesian estimator and Fisher information analysis, is given in Supplementary Information.

The trade-off between visibility and detection efficiency is further quantified through the Fisher information figure of merit $\mathcal{M} = \eta_{tot} V^2$ (Fig. 5b). The surface geometry illustrates the contrast between the linear scaling of information with efficiency and the quadratic sensitivity to visibility. Two representative designs are compared: detector A, a high-extinction design, and detector B, a high-absorption design. Despite lower visibility, detector B achieves a 46% higher Fisher information rate, allowing a $\sim$31% reduction in integration time required to reach a target state fidelity.

To confirm that all four pixels are necessary for complete tomography, we intentionally disable pixels one at a time and observe the effect on reconstruction fidelity (Fig. 5c). With one pixel removed (three detector configuration), fidelity plateaus near 83%, regardless of how many photons are accumulated. With two pixels removed, it plateaus near 73%. These ceilings are not set by noise; they are hard limits imposed by the dimensionality of the remaining measurement basis. A polarization qubit is described by three independent real parameters; once the instrument matrix no longer spans all three, the missing Stokes component becomes unobservable, and no amount of additional integration can recover it. The full four-pixel configuration has no blind spot, and its fidelity continues to improve with photon number until it converges to near unity.

The required photon budget to achieve 99% fidelity is quantified across three POVM geometries in Fig. 5d. To achieve near-perfect reconstruction with our basis ($\kappa = 3.23$), a factor of approximately 3.5 more photons are required compared to the theoretically optimal tetrahedral basis ($\kappa = 1.73$), but well within practical reach and far better than an arbitrary projection set ($\kappa = 14.15$), with all three cases confirming the expected $V^{-2}$ scaling.

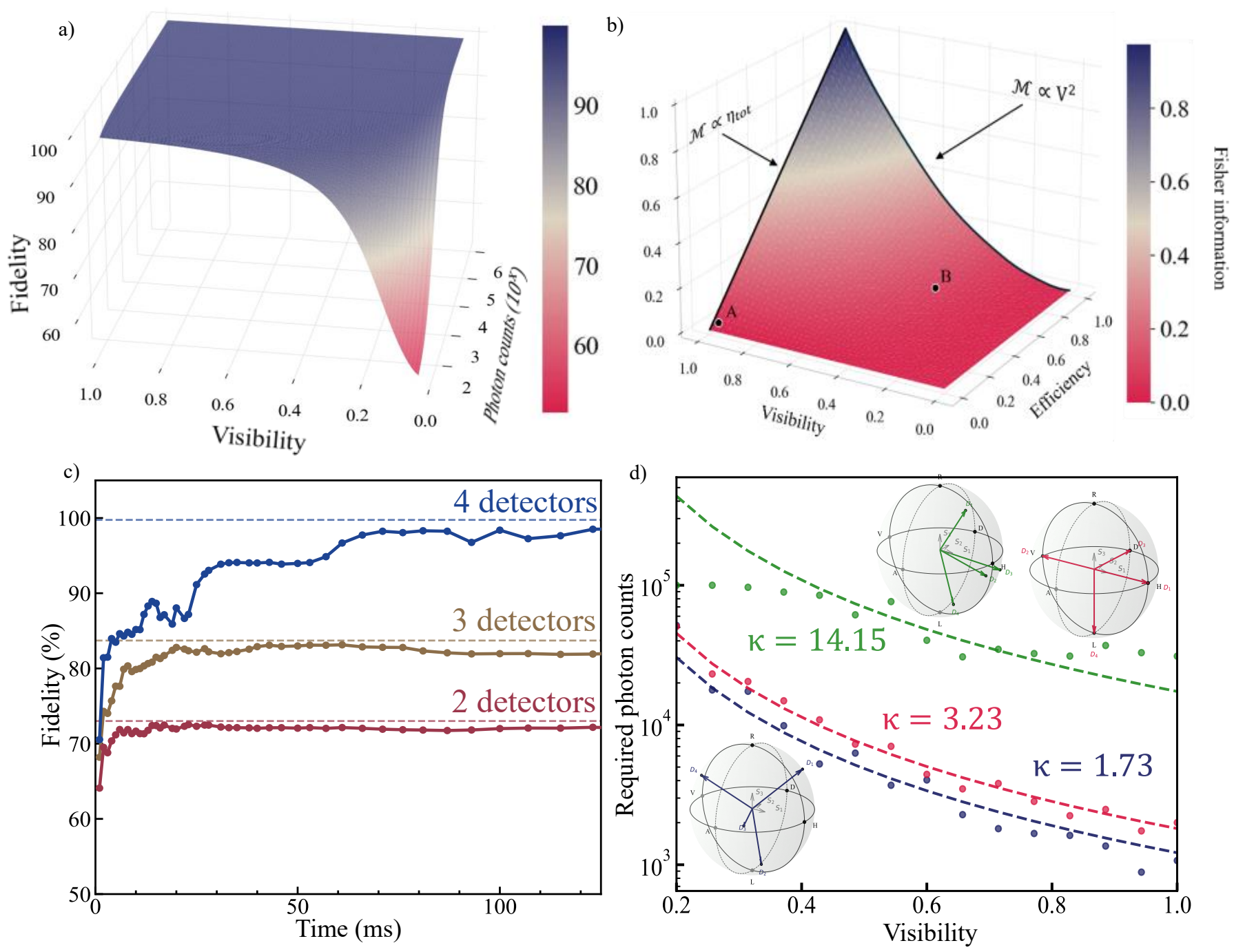

**Fig. 5. Theoretical limits of polarization state tomography with the M-SNSPD array.** (a) Analytical fidelity surface showing the transition from shot-noise to visibility-limited regimes. (b) Fisher information figure of merit $\mathcal{M} = \eta_{tot}V^2$ as a function of visibility and efficiency; points A and B compare a high-extinction and a high-absorption detector design, illustrating the quadratic penalty of reduced visibility and the non-linear compensation required to maintain a constant information rate. (c) Fidelity plateaus demonstrating the limits of incomplete measurements when using 2, 3 or 4 detectors. (d) Required photon budget for 99% fidelity across different POVM geometries, confirming $V^{-2}$ scaling and condition number penalties; insets show the corresponding measurement axis distributions on the Poincaré sphere.

## Discussion

We have demonstrated a hybrid metal-superconductor platform that bridges the gap between the functionality of optical metamaterials and the high-performance sensing of superconducting nanowires. By engineering the measurement operator directly upon absorption, we have moved beyond the "bucket-detector" paradigm, transforming SNSPDs from passive photon counters into active, state-resolved quantum measurement devices. Our four-pixel array demonstrates that complete, high-fidelity quantum state tomography can be achieved on-chip without the need for extrinsic, lossy, or mechanically unstable polarizing optics. While this demonstration focused on polarization, the underlying principle of near-field mode shaping is generalizable. Future iterations could focus on spectral, spatial or orbital angular momentum degrees of freedom [27], potentially enabling multi-mode detection on a single chip. Furthermore, the integration of a metallic overlayer provides a unique degree of freedom for electro-optic co-design. The ability to tune the metallic branch's impedance and thermal properties offers a path toward optimizing detector jitter and reset times without compromising optical selectivity. As integrated quantum photonic circuits continue to scale toward higher dimensionality and complexity [28], the ability to perform high-speed, on-chip state reconstruction will be important. This platform provides a robust framework for the next generation of multifunctional detectors in quantum communication, high resolution imaging and photonic quantum computing.

## Acknowledgments

Agency for Science, Technology and Research (QEP3-NQSP (H25-MNQ0931)).
National Centre for Advanced Integrated Photonics (NRF-MSG-2023-0002).

## Author contribution

P.B and J.W performed numerical simulations. J.W performed device nanofabrication. G.A performed optical spectroscopic characterization measurements. P.B performed single photon measurements. P.B., G.A, A.N.V. and C.S. analyzed the data and drafted the manuscript. C.S. supervised the work. All authors have accepted responsibility for the entire content of this manuscript and approved its submission.

## References

1. Kang, D., Kim, M., He, H. & Helmy, A. S. "Two polarization-entangled sources from the same semiconductor chip," Phys. Rev. A 92, 013821 (2015).
2. Duan, L.-M., Lukin, M. D., Cirac, J. I. & Zoller, P. "Long-distance quantum communication with atomic ensembles and linear optics," Nature 414, 413–418 (2001).
3. Knill, E., Laflamme, R. & Milburn, G. J. "A scheme for efficient quantum computation with linear optics," Nature 409, 46–52 (2001).
4. Bennett, C. H. & Brassard, G. "Quantum cryptography: Public key distribution and coin tossing," Theor. Comput. Sci. 560, 7–11 (2014; originally 1984).
5. Gol'tsman, G. N. et al. "Picosecond superconducting single-photon optical detector," Appl. Phys. Lett. 79, 705–707 (2001).
6. Marsili, F. et al. "Detecting single infrared photons with 93% system efficiency," Nat. Photonics 7, 210–214 (2013).
7. Semenov, A. D., Gol'tsman, G. N. & Korneev, A. A. "Quantum detection by current carrying superconducting film," Physica C 351, 349–356 (2000).
8. Meng, Y. et al. "Fractal superconducting nanowires detect infrared single photons with 84% system detection efficiency, 1.02 polarization sensitivity, and 20.8 ps timing resolution," ACS Photonics 9, 1547–1553 (2022).
9. Renema, J. J. et al. "Universal response curve for nanowire superconducting single-photon detectors," Phys. Rev. B 87, 174526 (2013).
10. Xiao, Y. et al. "Ultralow-filling-factor superconducting nanowire single-photon detector utilizing a 2D photonic crystal," Photon. Res. 11, 2128–2135 (2023).
11. Yun, Y. et al. "Superconducting-Nanowire Single-Photon Spectrometer Exploiting Cascaded Photonic Crystal Cavities," Phys. Rev. Appl. 13, 014061 (2020).
12. Toninelli, E., Ndagano, B., Vallés, A. et al. "Concepts in quantum state tomography and classical implementation with intense light: a tutorial," Adv. Opt. Photon. 11, 67–134 (2019).
13. Csete, M., Sipos, Á., Szalai, A. et al. "Improvement of infrared single-photon detectors absorptance by integrated plasmonic structures," Sci. Rep. 3, 2406 (2013).
14. Heath, R. M., Tanner, M. G., Drysdale, T. D. et al. "Nanoantenna Enhancement for Telecom-Wavelength Superconducting Single Photon Detectors," Nano Lett. 15, 819–822 (2015).
15. Novotny, L. & Hecht, B. "Surface plasmons," Principles of Nano-Optics, Cambridge University Press, 369–413 (2012).
16. Akhlaghi, M. K., Schelew, E. & Young, J. F. "Waveguide integrated superconducting single-photon detectors implemented as near-perfect absorbers of coherent radiation," Nat. Commun. 6, 8233 (2015).
17. Najafi, F. et al. "On-chip detection of non-classical light by scalable integration of single-photon detectors," Nat. Commun. 6, 5873 (2015).
18. Basiri, A., Chen, X., Bai, J. et al. "Nature-inspired chiral metasurfaces for circular polarization detection and full-Stokes polarimetric measurements," Light Sci. Appl. 8, 78 (2019).
19. Fedotov, V. A., Schwanecke, A. S., Zheludev, N. I., Khardikov, V. V. & Prosvirnin, S. L. "Asymmetric transmission of light and enantiomerically sensitive plasmon resonance in planar chiral nanostructures," *Nano Lett.* **7**, 1996–1999 (2007).

20. Gansel, J. K. et al. "Gold Helix Photonic Metamaterial as Broadband Circular Polarizer," Nano Lett. 10, 3596–3603 (2010).
21. Hentschel, M. et al. "Chiral plasmonics," Sci. Adv. 3, e1602735 (2017).
22. Esposito, M., Tasco, V., Todisco, F. et al. "Triple-helical nanowires by tomographic rotatory growth for chiral photonics," Nat. Commun. 6, 6484 (2015).
23. Hasan, S. B., Etrich, C., Filter, R., Rockstuhl, C. & Lederer, F. "Enhancing the nonlinear response of plasmonic nanowire antennas by engineering their terminations," Phys. Rev. B 88, 205125 (2013).
24. Miki, S., Yamashita, T., Terai, H. & Wang, Z. "High performance fiber-coupled NbTiN superconducting nanowire single photon detectors with Gifford-McMahon cryocooler," Opt. Express 21, 10208–10214 (2013).
25. Yang, J. K. W., Kerman, A. J., Dauler, E. A., Anant, V., Rosfjord, K. M. & Berggren, K. K. "Modeling the electrical and thermal response of superconducting nanowire single-photon detectors," IEEE Trans. Appl. Supercond. 17, 581–585 (2007).
26. Titchener, J. G., Gräfe, M., Heilmann, R. et al. "Scalable on-chip quantum state tomography," npj Quantum Inf. 4, 19 (2018).
27. Mair, A., Vaziri, A., Weihs, G. & Zeilinger, A. "Entanglement of the orbital angular momentum states of photons," Nature 412, 313–316 (2001).
28. Kimble, H. J. "The quantum internet," Nature 453, 1023–1030 (2008).